\documentclass[%
 aip,
 pof,%
 amsmath,amssymb,
reprint,%
]{revtex4-1}
\usepackage {CJK}
\usepackage{graphicx}
\usepackage{dcolumn}
\usepackage{bm}
\usepackage[mathlines]{lineno}
\usepackage{color}
\usepackage{hyperref}

\begin{document}
\begin{CJK*}{GB}{gbsn}

\title{Droplet splashing during the impact on liquid pools of shear-thinning fluids with yield stress}
\author{Xiaoyun Peng (彭小芸)}
\affiliation{State Key Laboratory of Engines, Tianjin University, Tianjin, 300072, China.}%
\author{Tianyou Wang (王天友)}
\affiliation{State Key Laboratory of Engines, Tianjin University, Tianjin, 300072, China.}%
\author{Kai Sun (孙凯)}
\affiliation{State Key Laboratory of Engines, Tianjin University, Tianjin, 300072, China.}%
\author{Zhizhao Che (车志钊)}
\email{chezhizhao@tju.edu.cn}
\affiliation{State Key Laboratory of Engines, Tianjin University, Tianjin, 300072, China.}

\date{\today}

\begin{abstract}
The impact of droplets on liquid pools is ubiquitous in nature and many industrial applications. Most previous studies of droplet impact focus on Newtonian fluids, while less attention has been paid to the impact dynamics of non-Newtonian droplets, even though non-Newtonian fluids are widely used in many applications. In this study, the splashing dynamics of shear-thinning droplets with yield stress are studied by combined experiments and simulations. The formation and the propagation of the ejecta sheet produced during the splashing process are considered, and the velocity, the radius, and the time of the ejecta sheet emergence are analyzed. The results show that the non-Newtonian fluid properties significantly affect the splashing process. The ejecta sheet of the splashing becomes easier to form as the flow index reduces, the large yield stress can affect the thickness of the ejecta sheet, and the spreading radius collapses into a geometrical radius due to that the inertia force is the dominant factor in the ejecta sheet propagation.
\end{abstract}


\maketitle
\end{CJK*}

\section{Introduction}\label{sec:sec01}
Droplet splashing during the impact on a solid or liquid surface occurs in many industrial and agricultural situations such as inkjet printing \cite{Basaran2013Inkjet, Ersoy2019DropletImpact}, coating \cite{Josserand2016DropImpact, Rein1993DropImpact}, fuel atomization \cite{Hammad2021Spay, Kavehpour2015DropImpact, Zhang2021DropletVapor, Ding2019DropImpact, Reitz1995DieselEngine, Yarin2006DropImpact}, cooling \cite{Wang2020DropImpact, Pasandideh-Fard2001DropImpact, Quere2013DropImpact}, and pesticide spraying \cite{Lin1998DropImpact, Spillman1984Spray}. The splashing process is featured by the generation of many small droplets upon the impact \cite{Worthington1882}. In the initial stage of the splashing, an ejecta sheet is produced at the intersection between the droplet and the air/liquid interface, and develops horizontally with a high speed \cite{Josserand2003DropletSplash, Weiss1999DropImpact}. Great progress has been in the experimental and theoretical study of the ejecta sheet \cite{Josserand2016DropletSplash, Agbaglah2015VortexAndJet, Zhang2012DropletSplash, Thoraval2012DropletSplash, Thoroddsen2002DropletSplash, Li2018EjectaSheet, Marcotte2019DropletSplash, Ren2020DropletSplash, Guo2018DropletSplash, Liu2021DropImpact, Liu2021DropletSplash, Liang2018DropletSplash, Deka2017DropletImpact, Guo2017DropletSplash}. The ejecta sheet originates from the liquid pool, not the droplet, as found in experiments \cite{Thoroddsen2002DropletSplash} and in simulations \cite{Josserand2016DropletSplash}. The shape of the ejecta sheet is the result of the interplay of several factors, including the inertial force, the surface tension, and the viscous force during the impact process \cite{Thoraval2012DropletSplash, Marcotte2019DropletSplash}. The high viscous force can rapidly decelerate the strong radial stretching of the ejecta sheet, and further makes the ejecta sheet bend outward like a bow \cite{Thoraval2012DropletSplash, Thoroddsen2002DropletSplash, Ren2020DropletSplash}. At high Reynolds numbers of impact ($\text{Re} \equiv {{{\rho }_{\text{L}}}D{{U}_{0}}}/{{{\mu }_{\text{L}}}}>2000$), the ejecta sheet appears in various shapes, and further breaks up into many tiny droplets; the vortex rings underneath the liquid surface are the reason for the instability of the ejecta sheet and the breakup \cite{Castrejon-Pita2012DropletSplash, Thoraval2012DropletSplash, Li2018EjectaSheet}. Further studies found that the vortex rings underneath the liquid surface also play an essential role in the formation of the ejecta sheet: the vortex rings detach from the surface, develop horizontally, and form a roll jet with no ejecta or a preceding ejecta \cite{Agbaglah2015VortexAndJet, Zhang2012DropletSplash, Li2018EjectaSheet}.

The impact of a non-Newtonian droplet on a solid substrate is a subject of many experiments and simulations such as for 3D printing \cite{Murphy20143DPrint, Luo20043DPrint}, and pesticides \cite{Bergeron2000NonnewtonDropletImpact} applications. It has been shown that a small amount of non-Newtonian solute can make remarkable changes in fluid properties, effectively controlling the flow behavior \cite{Bartolo2007NonnewtonDropletImpact, Bergeron2000NonnewtonDropletImpact, Thoraval2021NonnewtonDropletImpact}. For example, dilute aqueous solutions of a flexible polymer (e.g., polyethylene oxide) can effectively inhibit the process of retracting and rebounding when a droplet impacts on a hydrophobic surface without changing the viscosity of the solution greatly \cite{Bergeron2000NonnewtonDropletImpact}. The high elongational viscosity of the solution is considered as the reason for the retraction damping during the droplet impact \cite{Bergeron2000NonnewtonDropletImpact}. Further studies indicated that the rheological and surface properties are important factors for the spreading and retraction during the impact of non-Newtonian droplets \cite{Oishi2019NonnewtonDropletImpact, Bartolo2007NonnewtonDropletImpact, Luu2009NonnewtonDropletImpact, Guemas2012NonnewtonDropletImpact, Thoraval2021NonnewtonDropletImpact}. The rheological properties are always reflected by various physical models in simulation \cite{Oishi2019NonnewtonDropletImpact, Luu2009NonnewtonDropletImpact}. For example, the elastic effect is usually represented by an elastic spring, and the viscoplastic effect is represented by viscous elements \cite{Oishi2019NonnewtonDropletImpact, Luu2009NonnewtonDropletImpact}. For the impact of droplets with yield stress on smooth surfaces, the elasto-viscoplastic model \cite{Luu2009NonnewtonDropletImpact} can be used to capture the process of retracting, and this process has been analyzed theoretically, simulated numerically, and verified experimentally. A further study found that the process of spreading in experiments can be simulated better using an elasto-viscoplastic thixotropic model \cite{Oishi2019NonnewtonDropletImpact} than using the elasto-viscoplastic model \cite{Luu2009NonnewtonDropletImpact}, where the thixotropic features reflect the effect of viscosity with time. Besides the droplet rheology, the surface properties are also important for the spreading and retraction of non-Newtonian droplets. By changing the surface properties from hydrophilicity to super-hydrophobicity, the scaling law for the spreading in droplet impact is affected when $\text{We} \equiv  {{{\rho }_{\text{L}}}D{{U}_{0}}^{2}}/{\sigma }>1000$ \cite{Luu2009NonnewtonDropletImpact, Guemas2012NonnewtonDropletImpact}. For a super-hydrophobic surface, the inhibition effect by the elastic and shear-thinning effect to the retraction process during the impact of high-elastic droplets is not apparent \cite{Luu2009NonnewtonDropletImpact}. In addition, it has been found that the friction between droplet and substrate is more important than rheological properties in retraction inhibition effect \cite{Zang2014NonnewtonDropletImpact}. The friction force between the nanoparticles on the super-hydrophobic surface and the polymer aggregates in the droplet is the root of retraction damping during the impact \cite{Zang2014NonnewtonDropletImpact, Zang2013NonnewtonDropletImpact}. Even though the non-Newtonian properties can significantly alter the impact process, the studies mentioned above mainly focused on the retraction damping phenomenon in the droplet impact process, while the ejecta sheet of splashing process in non-Newtonian droplet impact has not been studied.

In this study, we focus on the ejecta sheet during the impact of shear-thinning droplets with yield stress on a liquid pool. Many yield-stress fluids have complex rheological properties such as viscoplastic force, yield stress, and even viscoelastic force at high polymer concentrations. Shear-thinning fluids with yield stress are a common type of non-Newtonian fluids, and are often observed in polymer solutions, molten polymers, complex fluids, and suspensions. The fluids exhibit shear-thinning behavior under high shear stress, and behave like a solid when the shear stress is less than a critical value. The yield stress can resist the plastic deformation in droplet impact processes. Since the formation of the ejecta sheet in the splashing process is essential to impact dynamics, we, in this study, compare the formation of the ejecta sheet for shear-thinning fluids with yield stress and Newtonian fluids in experimental measurements and numerical simulations. Then, the effects of the rheological parameters on the ejecta sheet are analyzed, including the velocity, the position, and the time of the ejecta sheet emergence.

\section{Experimental details }\label{sec:sec02}
\subsection{ Experimental setup }\label{sec:sec021}
\begin{figure}
  \centerline{\includegraphics[width=0.9\columnwidth]{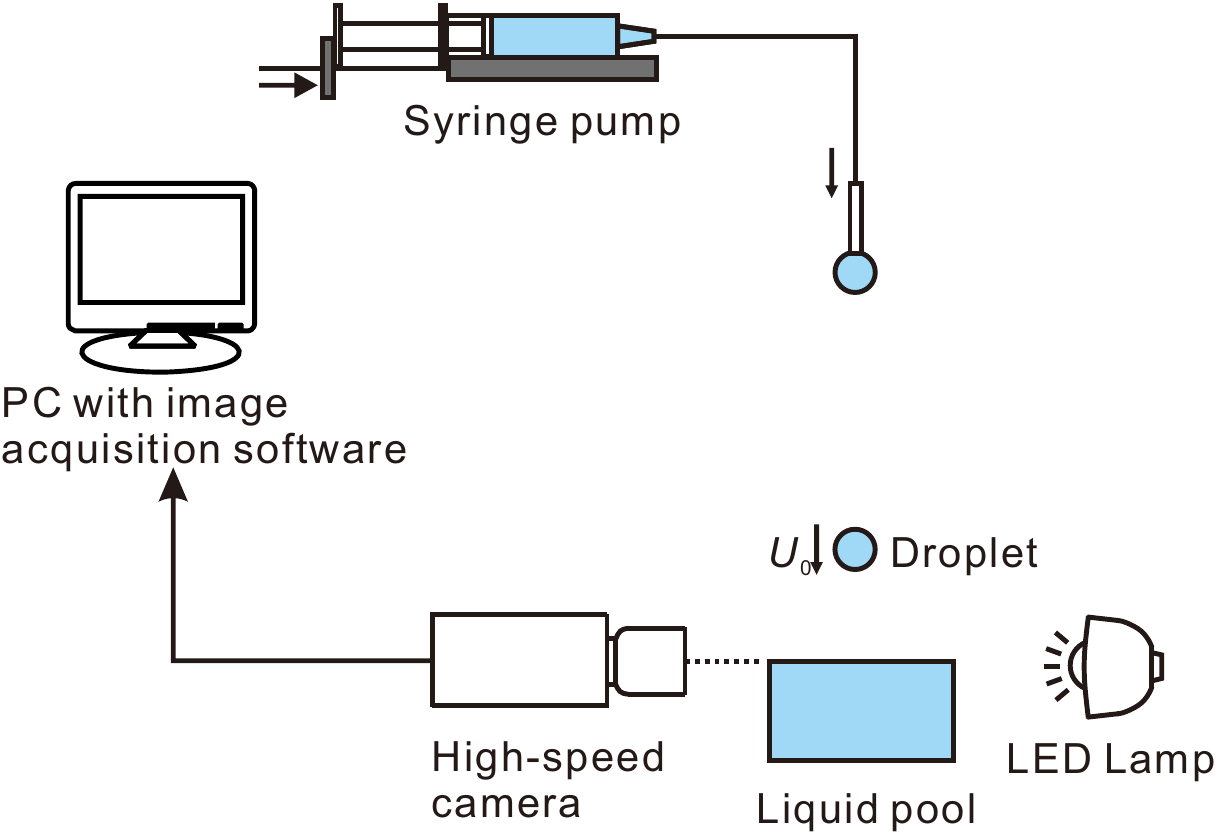}}%
  \caption{Schematic diagram of the experimental setup for the impact of droplets on liquid pools.}
\label{fig:fig01}
\end{figure}

The experimental setup is illustrated schematically in Fig.\ \ref{fig:fig01}. Droplets were produced at the tip of syringe needles, and their size was varied by changing the needle diameter. The syringe was pushed by a syringe pump (Harvard Apparatus, Pump 11 elite Pico plus) at a low speed to ensure the uniformity of droplet dripping. Then, droplets detached from the tip of the syringe needle, accelerated by gravity, and then fell into the liquid pool. We changed the initial speed of droplet impact by varying the falling height. A high-speed camera (Photron FASTCAM SA1.1) was used to record the droplet impact process at a frame rate of 16000 frames per second (fps) and a resolution of 512$\times$512 pixels. A high-power LED lamp (Hecho S5000, 60 W) was used to provide backlighting for high-speed imaging.

\subsection{ Fluid properties }\label{sec:sec022}
The non-Newtonian fluid used in this study is 0.05 wt\% Carbopol solution (Carbopol 940 supplied by Noveon), prepared using the same method as Luu and Forterre \cite{Luu2009NonnewtonDropletImpact}. First, Carbopol powder was slowly added into deionized (DI) water at 50 $^\circ$C, and continuously stirred at 500 rpm for several hours. Then, a sodium hydroxide solution of 18 wt\% was added into the solution to bring the pH from 3 up to 7. After that, the solution was adequately mixed at 700 rpm until no bubbles and lumps in the transparent gel.

\begin{figure}
  \centerline{\includegraphics[width=0.8\columnwidth]{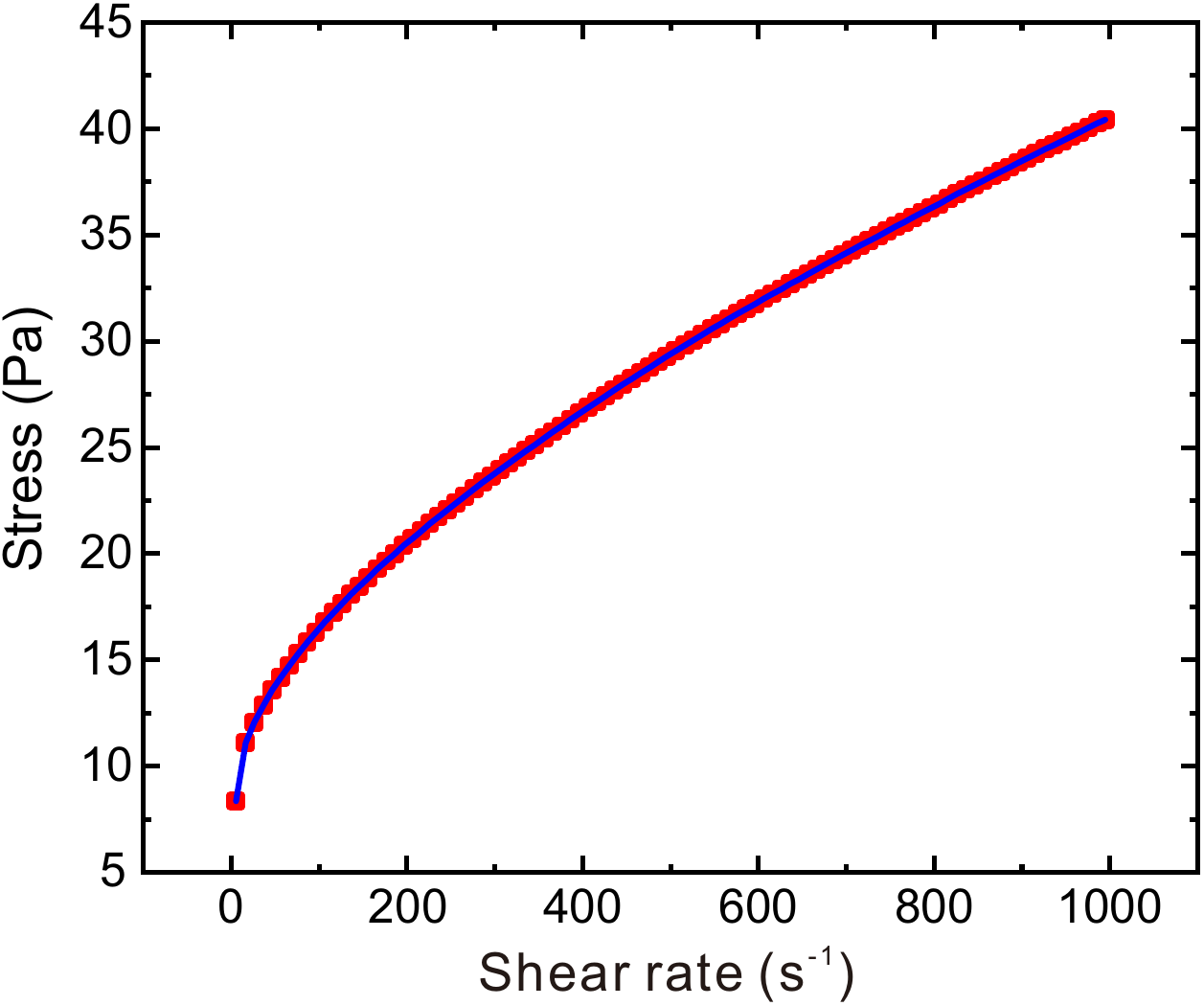}}%
  \caption{Steady-state stress versus shear rate for the 0.05 wt\% Carbopol solution. The red symbol represents the data measured by the rheometer, and the blue curve is fitting according to the Herschel--Bulkley model.}
\label{fig:fig02}
\end{figure}

The fluid properties of the Carbopol solution are listed in Tab.\ \ref{tab:tab1}. The surface tension of the Carbopol solution is approximately equal to that of pure water, as confirmed by surface-deformation spectroscopy \cite{Yoshitake2008DropletRheolgy}. Therefore, $\sigma = 0.07$ N/m was used as the surface tension of the Carbopol solution \cite{Luu2009NonnewtonDropletImpact}. The rheology of the fluids was measured using a rotational rheometer (TA Discovery HR-2) via steady-state shear measurements. In the steady-state shear measurements, the shear stress $\tau $ was measured as a function of the shear rate ${\dot{\gamma }= \partial u}/{\partial y}$ in the range of ${{10}^{-1}}$--${{10}^{3}}\ {{\text{s}}^{-1}}$, as shown in Fig.\ \ref{fig:fig02}. According to the previous studies \cite{Luu2009NonnewtonDropletImpact}, the Carbopol solution is a shear--thinning fluid with a yield stress, and this can be further verified by our experimental data shown in Fig.\ \ref{fig:fig02}, where the effective viscosity decreases as the shear rate increases. The relationship between the shear stress and the shear rate can be described well by the Herschel--Bulkley model $\tau ={{\tau }_{\text{c}}}+K{{\dot{\gamma }}^{n}}$, where ${{\tau }_{\text{c}}}$ is the yield stress, $K$ is the consistency index, and $n$ is the flow index \cite{Oishi2019NonnewtonDropletImpact, Luu2009NonnewtonDropletImpact, Saidi2010NonnewtonDropletImpact, Barnes1989NonnewtonRheology}. By fitting the measured data of shear stress and shear rate using the Herschel--Bulkley model, we can get the values of the rheological parameters of the non-Newtonian fluid (see Tab.\ \ref{tab:tab1}). The flow index $n$ and the consistency index $K$ are varied in the numerical simulation, and their ranges considered in this study are 0.5--1 and 0.05--0.8, respectively.

\begin{figure}
  \centerline{\includegraphics[width=0.9\columnwidth]{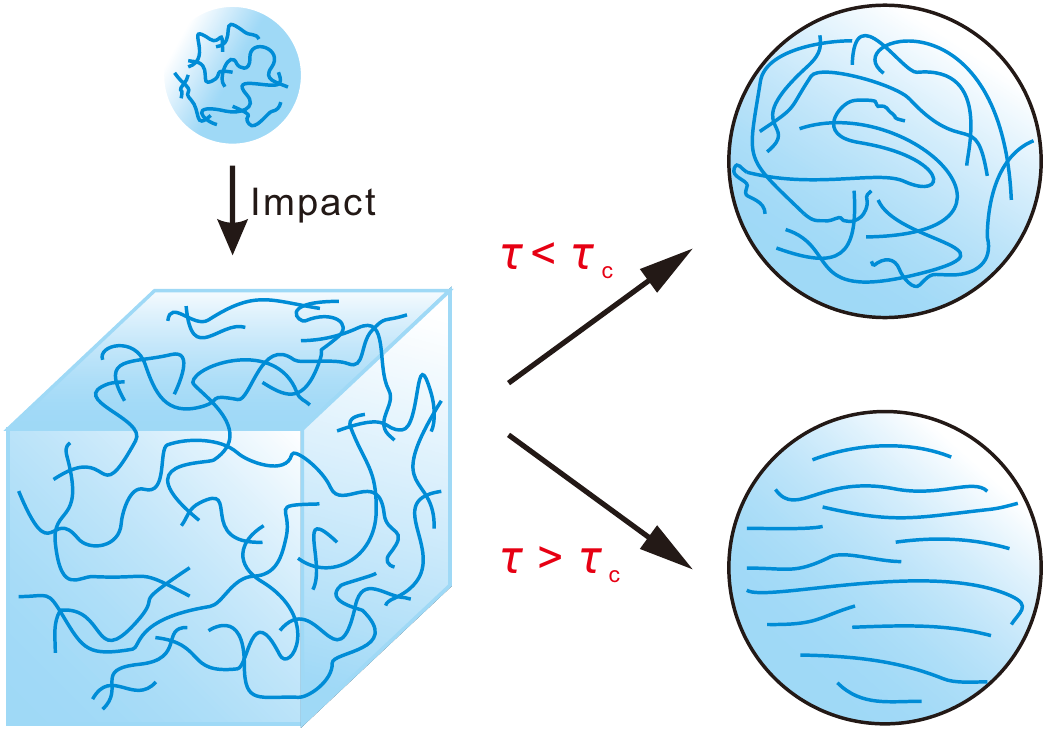}}%
  \caption{Schematic illustration of the structure of Carbopol gel.}
\label{fig:fig03}
\end{figure}

\begin{table*}[]
\renewcommand\arraystretch{1.3}
\begin{tabular}{p{6cm}p{3.5cm}p{3.5cm}p{3cm}}
\hline
Fluids                & Carbopol solution & Water+glycerol &Water+glycerol \\
                    &(0.05 wt\%) &(84 wt\%) &(85 wt\%) \\
\hline
Yield stress, ${{\tau }_{c}}\;(\text{Pa})$                     & 8.239 $\pm$ 0.116          & --      &  --     \\
Consistency index, $K\;(\text{Pa}\cdot {{\text{s}}^{\text{n}}})$                      & 0.505 $\pm$ 0.017         & --      & --      \\
Flow index, $n$                      & 0.601 $\pm$ 0.004         & --      &-- \\
Dynamic viscosity, ${{\mu }_{\text{L}}}\;(\text{Pa}\cdot \text{s})$                      & --         & $5.368\times {{10}^{-2}}$      & $6.005\times {{10}^{-2}}$\\
Density, ${{\rho }_{\text{L}}}\;(\text{kg}/{{\text{m}}^{3}})$                     & 1010 & 1217 & 1220\\
Surface tension, $\sigma \;(\text{N}/{{\text{m}}})$                     & 0.07         &0.0654       &0.0654 \\
\hline
\end{tabular}
\caption{Fluid properties in the droplet impact experiment.}\label{tab:tab1}
\end{table*}

The non-Newtonian rheological properties of the Carbopol gel originate from the internal molecular structure of the solution. As illustrated in Fig.\ \ref{fig:fig03}, the Carbopol powders swell with neutralization, and a cross-linked molecular structure is formed even at a low polymer concentration \cite{Kim2003ChemistryCarbopol, Shafiei2018ChemistryCarbopol}. As a result, when a droplet of the Carbopol solution impacts on the pool of the same liquid, the molecules are entangled and block the liquid flow. The polymers remain entangled in the liquid at a low shear rate, and the fluid shows a high effective viscosity. However, at an increased shear rate, the effective viscosity of the fluid decreases dramatically. This is because the cross-linked molecular structure of fluid is destroyed when the yield stress of the fluid is overcome. At last, the cross-linked structure completely breaks with the polymer untangling completely. Therefore, the effective viscosity decreases as the shear rate increases, producing the shear-thinning property.

To quantify the effects of the non-Newtonian rheological properties on the impact process, the following dimensionless numbers are used in the experiments and the simulations to analyze the dynamics of droplet impact. The Reynolds number Re is used to indicate the ratio between the inertial force and the viscous force
\begin{equation}\label{eq:eq01}
\text{Re} \equiv \frac{{{\rho }_{\text{L}}}D{{U}_{0}}}{K{{\left( {{{U}_{0}}}/{D} \right)}^{n-1}}}=\frac{{{\rho }_{\text{L}}}{{D}^{n}}{{U}_{0}}^{2-n}}{K},
\end{equation}
where ${{\rho }_{\text{L}}}$ is the density of the fluid, $D$ is the initial diameter of the droplet, and ${{U}_{0}}$ is the initial velocity of droplet impact. For the Newtonian fluids (i.e., $n=1$ and $K={{\mu }_{\text{L}}}$), the definition of the Reynolds number recovers the original definition $\text{Re} \equiv {{{\rho }_{\text{L}}}D{{U}_{0}}}/{{{\mu }_{\text{L}}}}$.

It should be noted that both $K$ and $n$ affect the Reynolds number. Therefore, in some simulations, to maintain the same Reynolds number while changing $n$, we change $K$ accordingly to make sure $\text{Re}$ is unchanged. In addition, the Oldroyd number $\text{Od}$ is used to compare the yield stress with the viscous effect
\begin{equation}\label{eq:eq02}
  \text{Od} \equiv \frac{{{\tau }_{\text{c}}}}{K{{\left( {{{U}_{0}}}/{D} \right)}^{n}}},
\end{equation}
where ${{\tau }_{\text{c}}}$ is the yield stress. For Newtonian fluids, the Oldroyd number is zero as ${{\tau }_{\text{c}}}=0$. The Od number in the simulations is varied by changing the yield stress (${{\tau }_{\text{c}}}$) while fixing the characteristic shear rate (${\dot{\gamma }={{U}_{0}}}/{D}$) and the consistency index ($K$).

The Weber number $\text{We}$ is used to indicate the ratio between the inertial force and the surface tension force
\begin{equation}\label{eq:eq03}
  \text{We} \equiv  \frac{{{\rho }_{\text{L}}}D{{U}_{0}}^{2}}{\sigma },
\end{equation}
where $\sigma $ represents the surface tension of the fluid.

The Capillary number $\text{Ca}$ is used to indicate the ratio between the viscous force and the surface tension force
\begin{equation}\label{eq:eq02}
  \text{Ca} \equiv \frac{\text{We}}{\text{Re}}=\frac{K{{\left( {{{U}_{0}}}/{D} \right)}^{n-1}}{{U}_{0}}}{\sigma }.
\end{equation}
The above dimensionless parameters considered in this study are summarized in Tab.\ \ref{tab:tab2}.

The results of the non-Newtonian fluids are compared with that of Newtonian fluids. The Newtonian fluids are selected based on that they have the same characteristic viscosity, i.e., the viscosity of the non-Newtonian fluids at the characteristic shear rate ${{{U}_{0}}}/{D}$. We used glycerol solutions of 84 wt\% and 85 wt\% at 30 $^\circ$C as the experimental Newtonian fluid. The viscosity of the 84 wt\% and 85 wt\% glycerol solutions are ${{\mu }_{\text{L}}}=5.368\times {{10}^{-2}}\ \text{Pa}\cdot \text{s}$ and ${{\mu }_{\text{L}}}=6.005\times {{10}^{-2}}\ \text{Pa}\cdot \text{s}$, respectively, which are equal to the characteristic viscosity of 0.05 wt\% Carbopol solution with an impact velocity of ${{U}_{0}}=1.26\ \text{m/s}$ to ${{U}_{0}}=2.10\ \text{m/s}$. Meanwhile, the surface tension of the glycerol solutions is nearly not changed with the viscosity. It can be ensured that there is no interference from other variables when comparing the droplet impact process with the same characteristic viscosity.

\begin{table*}[]
\renewcommand\arraystretch{1.8}
\begin{tabular}{p{6cm}p{6cm}p{4cm}}
\hline
Dimensionless numbers & Definitions & Ranges \\
\hline
Reynolds number    & $\text{Re} \equiv \frac{{{\rho }_{\text{L}}}D{{U}_{0}}}{K{{\left( {{{U}_{0}}}/{D} \right)}^{n-1}}}$     & 91.05--3366.7 \\
Weber number      & $\text{We} \equiv  \frac{{{\rho }_{\text{L}}}D{{U}_{0}}^{2}}{\sigma }$    & 72.14--242.48 \\
Oldroyd number      & $\text{Od} \equiv \frac{{{\tau }_{\text{c}}}}{K{{\left( {{{U}_{0}}}/{D} \right)}^{n}}}$     & 0--6.72 \\
Capillary number    & $\text{Ca} \equiv \frac{\text{We}}{\text{Re}}=\frac{K{{\left( {{{U}_{0}}}/{D} \right)}^{n-1}}{{U}_{0}}}{\sigma }$    & 0.02--2.66 \\
\hline
\end{tabular}
\caption{Dimensionless parameters in the droplet impact simulation.}\label{tab:tab2}
\end{table*}

\section{ Numerical methods }\label{sec:sec03}

\begin{figure}
  \centerline{\includegraphics[width=0.8\columnwidth]{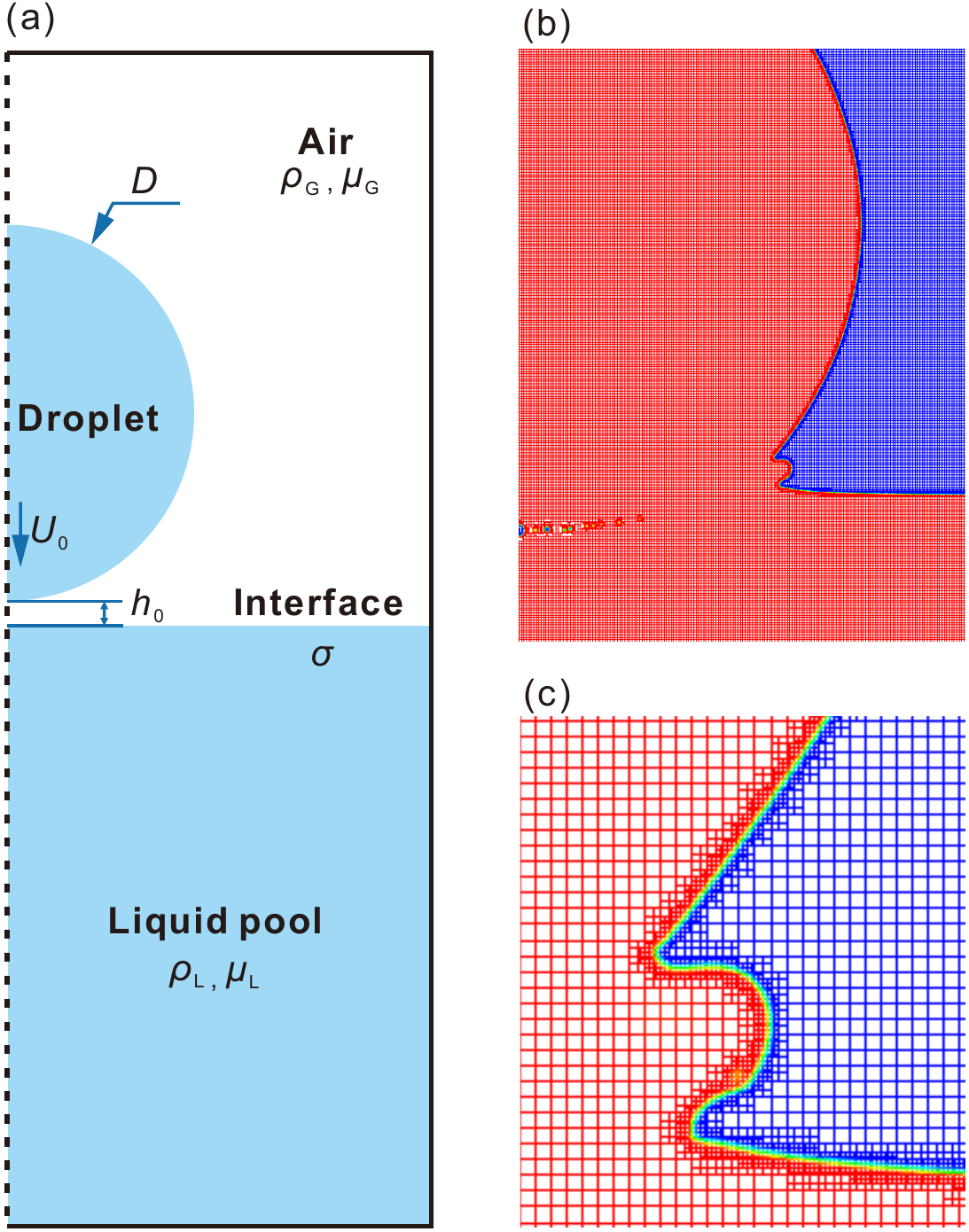}}%
  \caption{Schematic diagram of the numerical simulation and the grid refinement. Here, (a) represents a spherical droplet of diameter $D$ with a velocity of ${{U}_{0}}$ impacting on the liquid pool at a distance ${{h}_{0}}$. Adaptive grid refinement is used as shown in (b), and an enlarged drawing of the mesh is shown in (c).}
\label{fig:fig04}
\end{figure}

The simulation is performed in a 2D axisymmetric domain, as shown in Fig.\ \ref{fig:fig04}(a). We considered a droplet of diameter $D$, density ${{\rho }_{\text{L}}}$, and dynamic viscosity ${{\mu }_{\text{L}}}$ impacting at a speed of ${{U}_{0}}$ on a liquid pool of the same liquid. The gas had a density of ${{\rho }_{\text{G}}}$ and a dynamic viscosity of ${{\mu }_{\text{G}}}$, and the surface tension of the liquid interface was $\sigma $. As shown in Fig.\ \ref{fig:fig04}(a), we started the simulations with a small air gap ${{h}_{0}}$ between the droplet and the surface of the liquid pool. The small gap can significantly reduce the simulation time without affecting the accuracy of the calculation. It has been demonstrated that if the ratio between the air gap and the droplet diameter ${{{h}_{0}}}/{D}$ is larger than 1/30, the results of the impact dynamics are not affected \cite{Josserand2016DropletSplash}.

The fluids used in our study are all incompressible, with constant fluid properties, e.g., the density and the surface tension. We used the axisymmetric incompressible Navier-Stokes equations with the one-fluid formulation
\begin{equation}\label{eq:eq05}
  \nabla \cdot \mathbf{u}\text{=0},
\end{equation}
\begin{equation}\label{eq:eq06}
  	\rho \left( \frac{\partial \mathbf{u}}{\partial t}+\mathbf{u}\cdot \nabla \mathbf{u} \right)=-\nabla p+\nabla \cdot \mu \left[ \nabla \mathbf{u}+{{\left( \nabla \mathbf{u} \right)}^{\text{T}}} \right]+\sigma \kappa {{\delta }_{s}}\mathbf{n},
\end{equation}
where $\mathbf{u}$ is the flow velocity, $\rho $ is the fluid density, $\mu $ is the fluid viscosity, and $p$ is the pressure. Moreover, $\mathbf{n}$ is the unit vector normal to the interface, $\kappa $ is the curvature of the interface, and ${{\delta }_{s}}$ is the Dirac distribution and indicates that the surface tension effect is concentrated at the interface.

We used the multiphase flow solver interDyMFoam in OpenFOAM for the simulation. The solver was based on the volume of fluid (VOF) method to capture the interface, which uses a volume fraction function $\alpha$ of the interest phase in a computational grid cell. When a cell is empty, the value of $\alpha$ is zero; when a cell has the traced fluid inside but not full, $0<\alpha<1$; and when a cell is full, $\alpha=1$. Therefore, the VOF equation is
\begin{equation}\label{eq:eq07}
  \frac{\partial \alpha }{\partial t}+\mathbf{u}\cdot \nabla \alpha =0.
\end{equation}

The grid used in the simulation is shown in Fig.\ \ref{fig:fig04}b. The base grid for the simulation was $240\times 840$, and an adaptive mesh refinement (AMR) at the interface was used to improve the simulation accuracy. With two levels of refinement, the smallest cell had a size of $\Delta x=3.125\; \mu \text{m}$, as shown in Fig.\ \ref{fig:fig04}(c). To capture the flow details with reasonable computational resources, we did a mesh-independency study by changing the smallest cell from ${D}/250$, ${D}/500$, ${D}/1000$ to ${D}/2000$, and also compared the numerical simulations with the experimental images for the development of the ejecta sheet. As shown in Fig.\ S1 in Supplementary Material, the mesh with the smallest cell ${D}/1000$ is enough to capture the ejecta sheet. Hence, this mesh density was used for further simulations.

\section{ Results and Discussion }\label{sec:sec04}
\subsection{ Comparison between experiments and simulations }\label{sec:sec041}

\begin{figure}
  \centerline{\includegraphics[width=1\columnwidth]{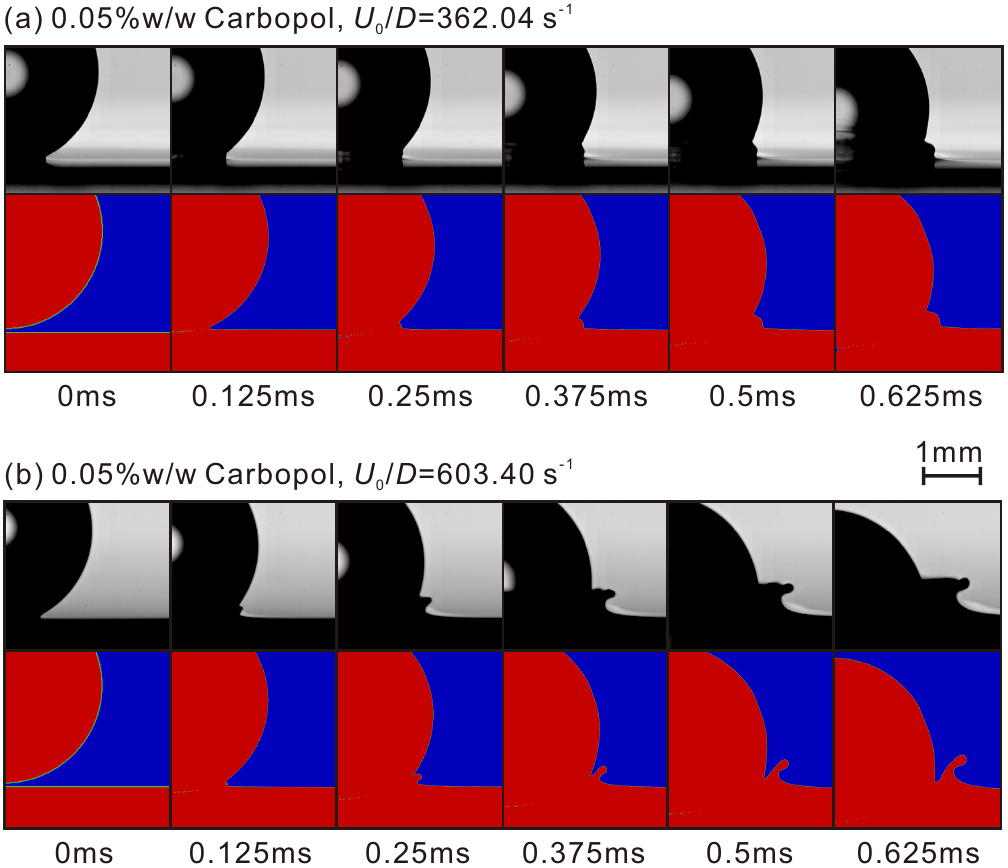}}%
  \caption{Comparison between the numerical simulations and the experimental images for the development of the ejecta sheet upon the impact of non-Newtonian droplets. The droplet is 0.05 wt\% Carbopol solution with $D=3.47\ \text{mm}$. The impact speeds are (a) $U_0=1.26\ {\text{m/s}}$ and (b) $U_0=2.10\ \text{m/s}$, respectively. Multimedia views: Movies 1; Movies 2. }
\label{fig:fig05}
\end{figure}

A comparison of the impact process between the experiments and the simulations is made to validate the numerical model. Fig.\ \ref{fig:fig05}(a) (Multimedia views) shows the impact process of a Carbopol droplet with $D=3.47\ \text{mm}$ and $U_0=1.26\ \text{m/s}$, while Fig.\ \ref{fig:fig05}(b) (Multimedia views) shows the impact process at a higher velocity $U_0=2.10\ \text{m/s}$. We can see that the simulation agrees with the experiment well in both cases. The simulation can also capture the ejecta sheet precisely, as shown in Fig.\ \ref{fig:fig05}(b) (Multimedia views). The ejecta sheet appears from the contact interface between the droplet and the liquid pool, and develops upward and outward quickly. The comparison indicates the numerical model used in the simulation has high accuracy and can successfully capture the impact dynamics of the Carbopol droplet.

\begin{figure}
  \centerline{\includegraphics[width=1\columnwidth]{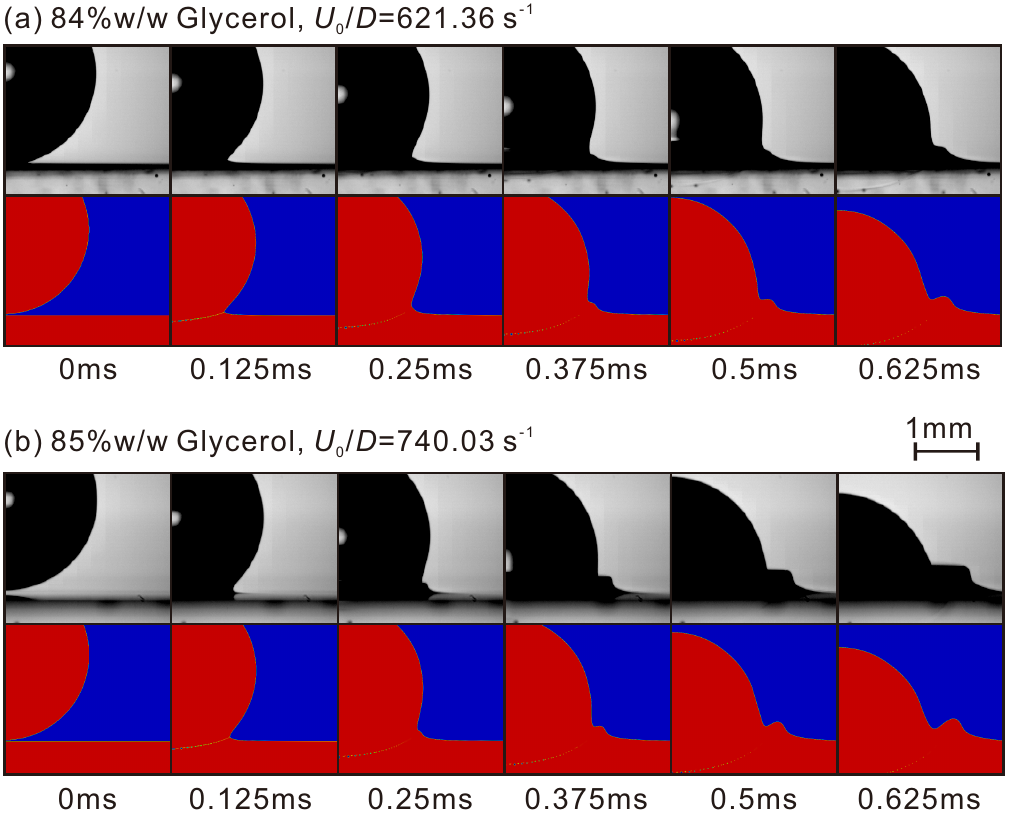}}%
  \caption{Comparison between the numerical simulations and the experimental images for the development of the ejecta sheet upon the impact of Newtonian droplets. (a) 84 wt\% glycerol solution, $D=2.92\ \text{mm}$, ${{U}_{0}}=1.81 \text{m/s}$; (b) 85 wt\% glycerol solution, $D=2.96\ \text{mm}$, ${{U}_{0}}=2.19\ \text{m/s}$. Multimedia views: Movies 3; Movies 4. }
\label{fig:fig06}
\end{figure}

To further verify the accuracy of the numerical simulation, we also compare the impact dynamics of Newtonian droplets between experiments and simulations. Glycerol solutions with the same characteristic viscosity as that of the Carbopol solution are used. The droplet in Fig.\ \ref{fig:fig06}(a) (Multimedia views) is 84 wt\% glycerol solution with $D=2.92\ \text{mm}$, ${{U}_{0}}=1.81\ \text{m/s}$, and ${{\mu }_{\text{L}}}=5.368\times {{10}^{-2}}\ \text{Pa}\cdot \text{s}$, while the droplet in Fig.\ \ref{fig:fig06}(b) (Multimedia views) is 85 wt\% glycerol solution with $D=2.96\ \text{mm}$, ${{U}_{0}}=2.19\ \text{m/s}$, and ${{\mu }_{\text{L}}}=6.005\times {{10}^{-2}}\ \text{Pa}\cdot \text{s}$. We can see that the simulations and experiments agree well. The comparison verifies that the simulations are also suitable for calculating the impact dynamics of Newtonian droplets.

\subsection{ Comparison between Newtonian and non-Newtonian fluids }\label{sec:sec042}

\begin{figure*}
  \centerline{\includegraphics[width=1.3\columnwidth]{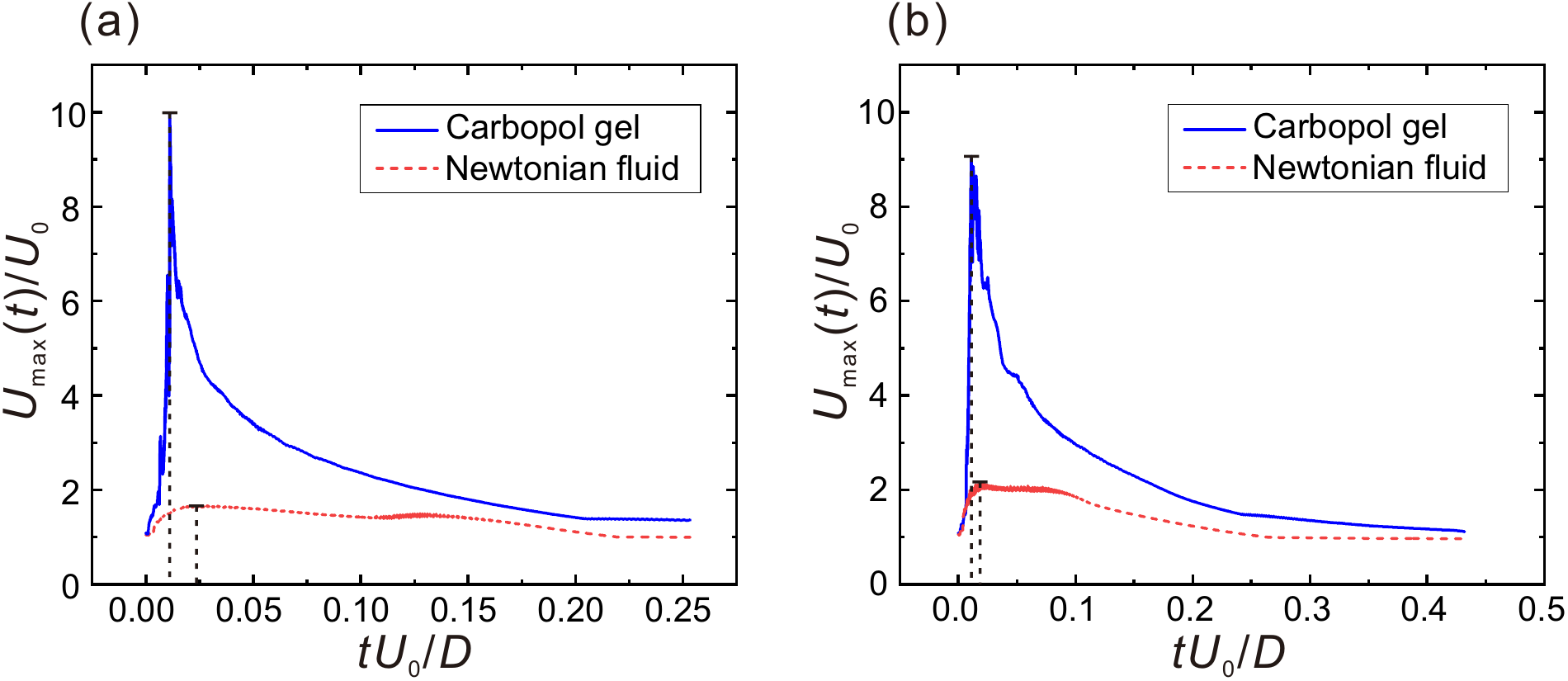}}%
  \caption{Comparison of the dimensionless maximum velocity ${{{U}_{\text{max}}}}\left( t \right)/{{{U}_{0}}}$ as a function of the dimensionless time $t{{{U}_{0}}}/{D}$ between the Carbopol gel and the Newtonian fluid, (a) for $\text{Re}=92.31$ and $\text{We}=79.54$, and (b) for $\text{Re}=188.72$ and $\text{We}=220.94$. The blue solid curves in Figs.\ \ref{fig:fig07}(a) and \ref{fig:fig07}(b) correspond to the non-Newtonian cases shown in Figs.\ \ref{fig:fig05}(a) and \ref{fig:fig05}(b), respectively, and the red dashed curves correspond to the Newtonian droplet at the same Reynolds numbers.}
\label{fig:fig07}
\end{figure*}

To quantitatively compare the impact dynamics between the non-Newtonian droplet and the Newtonian droplet, the formation of the ejecta sheet is considered for Newtonian and non-Newtonian droplets at the same characteristic shear rate, i.e., the same Reynolds number. We define the instant of the spherical droplet contacting the liquid surface as $t=0$, and define the dimensionless time as $t{{{U}_{0}}}/{D}$. Fig.\ \ref{fig:fig07} shows the evolution of the dimensionless maximum velocity ${{{U}_{\text{max}}}}\left( t \right)/{{{U}_{0}}}$, where ${{U}_{\text{max}}}\left( t \right)$ is measured in the simulation by finding the maximum velocity in the liquid above the initial surface plane of the liquid pool (to eliminate the velocity interference caused by bubble collapse and vortex below the surface of the liquid pool, See Fig.\ S2 in Supplementary Material). Therefore, the peak in the maximum velocity curve occurs when the interface curvature reverses (i.e., the time is ${{t}_{\text{j}}}$), corresponding to the emergence of the ejecta sheet \cite{Josserand2016DropletSplash}, which can be verified by the velocity field. As shown in Fig.\ \ref{fig:fig07}(a) and \ref{fig:fig07}(b), the magnitude of the peak velocity for the non-Newtonian fluid is much higher than that of the Newtonian fluid. In addition, the emergence time of the ejecta sheet for the non-Newtonian droplet is ahead of that for the Newtonian fluid. This result indicates that the ejecta sheet for the non-Newtonian droplet is faster and earlier than that of the Newtonian droplet at the same Reynolds number. This comparison also demonstrates that the rheological properties have a profound effect on the impact dynamics and the splashing process.

The large emerging velocity of the ejecta sheet for the non-Newtonian fluid shown in Fig.\ \ref{fig:fig07} can be explained from the shear-thinning property of the droplet. At the instant of ejecta sheet emergence, a large radial velocity is produced at the contact point by the downward movement of the droplet fluid according to the mass conservation. Therefore, a large shear rate is formed locally, which corresponds to a low effective viscosity according to the shear-thinning property (see Fig.\ \ref{fig:fig02}). The local effective viscosity at the point of ejecta sheet emergence is much smaller than that of the corresponding Newtonian droplet. As a consequence, the ejecta sheet is easier to form for the shear-thinning droplet, and the velocity of the ejecta sheet emergence for the shear-thinning droplet is much larger than that of the Newtonian droplet.

\begin{figure}
  \centerline{\includegraphics[width=0.8\columnwidth]{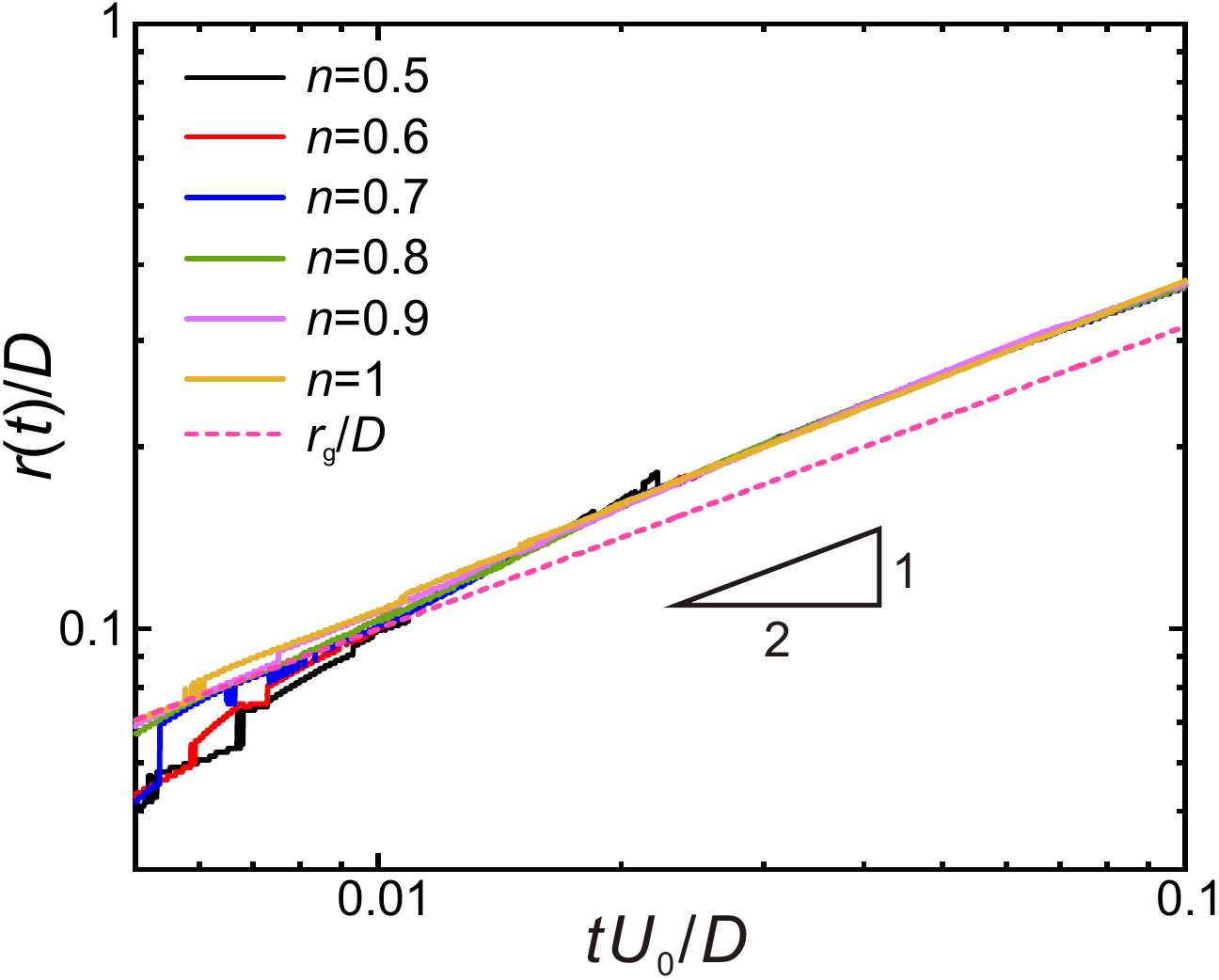}}%
  \caption{Dimensionless spreading radius ${r}\left( t \right)/{D}$ as a function of the dimensionless time $t{{{U}_{0}}}/{D}$ in a log-log plot for various $n$ values. The Reynolds number is $\text{Re}=198.69$. The Weber number is $\text{We}=219.91$. The straight line represents the geometrical law ${{{r}_{\text{g}}}}\left( t \right)/{D}={{\left( {t{{U}_{0}}}/{D} \right)}^{{1}/{2}}}$.}
\label{fig:fig08}
\end{figure}

The effect of the rheological parameters on the propagation of the ejecta sheet is analyzed. To quantitatively describe the development of the ejecta sheet, the spreading radius of the impact ${r}\left( t \right)$ is used and is defined as the horizontal scale from the axis to the point in the liquid where the velocity is maximal \cite{Josserand2016DropletSplash, Josserand2003DropletSplash} (See Fig.\ S2 in Supplementary Material). The dimensionless spreading radius ${r}\left( t \right)/{D}$ as a function of the dimensionless time $t{{{U}_{0}}}/{D}$ for different values of $n$ is plotted in the logarithm scale in Fig.\ \ref{fig:fig08}. The dimensionless spreading radius ${r}\left( t \right)/{D}$ increases as the ejecta sheet develops, and collapses into a single curve for different values of $n$. In the study of the splashing process by Josserand et al.\ \cite{Josserand2016DropletSplash, Josserand2003DropletSplash}, a geometrical law for the spreading radius ${r}\left( t \right)$ was proposed based on the mass conservation, ${r}\left( t \right)\simeq {{r}_{\text{g}}}\left( t \right)=\sqrt{D{{U}_{0}}t}$ (See Fig.\ S3 in Supplementary Material for a schematic diagram). In the model, the inertial force during the impact process is considered, and the viscous dissipation and the droplet deformation are neglected. The geometrical law \cite{Josserand2016DropletSplash, Josserand2003DropletSplash} can be written in the dimensionless form as ${{r}_{\text{g}}}\left( t \right)/{D}={{\left( {t{{U}_{0}}}/{D} \right)}^{{1}/{2}}}$, which is a straight line with a slope of 1/2, as shown in Fig.\ \ref{fig:fig08}. For the dimensionless spreading radius ${r}\left( t \right)/{D}$ of our simulation, all the curves collapse into a single curve that is almost parallel to the geometrical law. This result means that the non-Newtonian characteristics in our study do not affect the spreading radius. This is because the inertial force is still the dominant factor for the development of the ejecta sheet in non-Newtonian droplet impact, and the viscous dissipation is still negligible.

The results in Fig.\ \ref{fig:fig07} and Fig.\ \ref{fig:fig08} also show that the effects of the shear-thinning properties on the ${{{U}_{\text{max}}}}\left( t \right)$ and on the spreading radius ${r}\left( t \right)$ are different. The shear-thinning properties have a strong effect on ${{{U}_{\text{max}}}}\left( t \right)$, but its effect on the spreading radius is insignificant. Since the local velocity near ${{{U}_{\text{max}}}}\left( t \right)$ is significantly larger than the mean velocity, a small region of high speeds is produced locally, and results in a large velocity gradient locally. The local large velocity gradient produces a viscous boundary layer (which later determines the formation of the ejecta sheet \cite{Josserand2016DropletSplash}. Therefore, the maximum velocity ${{{U}_{\text{max}}}}\left( t \right)$ is significantly affected by the shear-thinning properties, as shown in Fig.\ \ref{fig:fig07}. In contrast, the flow in other regions does not have a large velocity gradient. Therefore, the viscous effect is negligible, and the flow is dominated by the inertia force.

\subsection{ Effect of $n$ }\label{sec:sec043}

\begin{figure*}
  \centerline{\includegraphics[width=1.7\columnwidth]{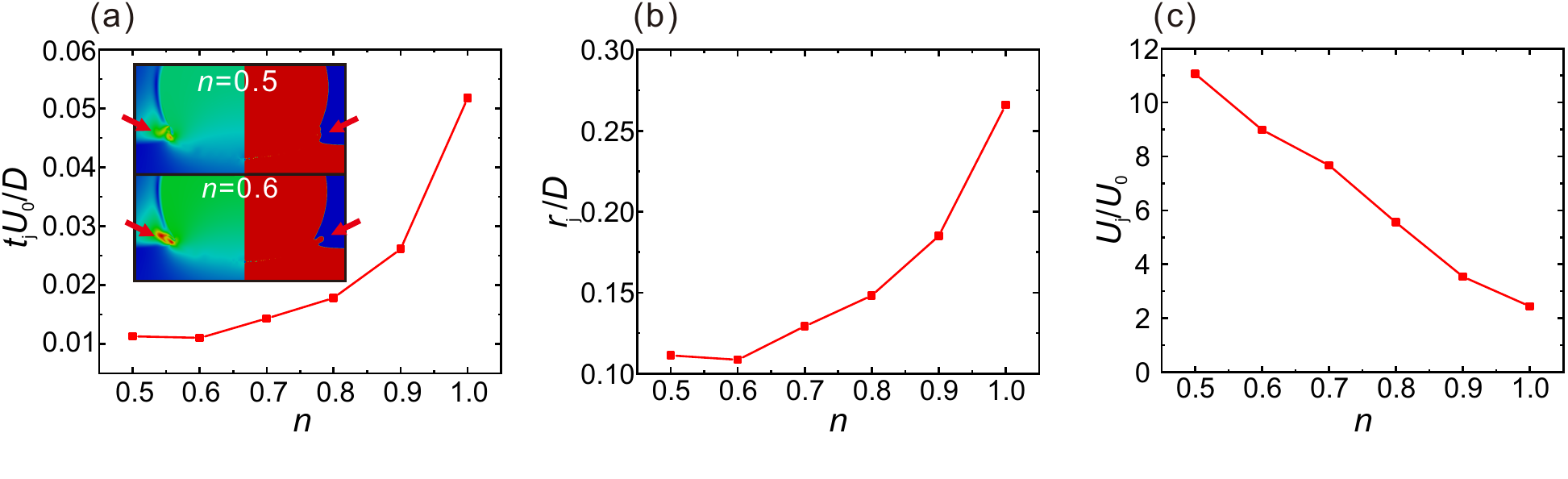}}%
  \caption{(a) The dimensionless time of ejecta sheet emergence ${{{t}_{\text{j}}}{{U}_{0}}}/{D}$, (b) the dimensionless spreading radius of ejecta sheet emergence ${{{r}_{\text{j}}}}/{D}$, (c) the dimensionless velocity of ejecta sheet emergence ${{{U}_{\text{j}}}}/{{{U}_{0}}}$ as a function of $n$. The Reynolds number is $\text{Re}=198.69$. The Weber number is $\text{We}=219.91$. The two inset figures in Fig.\ \ref{fig:fig09}(a) show the jet of $n=0.5$ and $n=0.6$, and the left halves of the inset figures are the velocity flow fields of the impact processes, while the right halves are the interface shapes. The roll jet and the ejecta sheet are highlighted by arrows for $n=0.5$ and $n=0.6$, respectively.}
\label{fig:fig09}
\end{figure*}

The influence of rheological parameters on the ejecta sheet during droplet impact is analyzed by varying $n$. The Reynolds number is fixed to ensure that the viscosity effect is comparable as $n$ varies in the simulations. As shown in Fig.\ \ref{fig:fig09}, $n$ affects the time, the radius, and the velocity of ejecta sheet emergence. As $n$ decreases from 1, the time of the ejecta sheet emergence becomes smaller, the spreading radius of ejecta sheet emergence becomes smaller, and the velocity of ejecta sheet emergence increases. This effect can also be explained by the shear-thinning property of the fluid. The local effective viscosity at the point of the ejecta sheet emergence is much smaller for shear-thinning droplets than that for the Newtonian droplets, as explained in Sec.\ \ref{sec:sec042}. A smaller value of $n$ indicates a stronger shear-thinning effect. As $n$ decreases, the local effective viscosity decreases due to the high local velocity and the high local shear rate. Therefore, the local viscous dissipation decreases, and the ejecta sheet becomes easier to form. Therefore, as $n$ decreases, the time and the spreading radius of the ejecta sheet emergence become smaller, and the velocity of the ejecta sheet emergence becomes larger.

From Figs.\ \ref{fig:fig09}a and \ref{fig:fig09}b, it can also be seen that $n=0.5$ is a turning point for the variation in the time and the spreading radius of ejecta sheet emergence. The two inset images in Fig.\ \ref{fig:fig09}(a) show the snapshots at $n=0.5$ and $n=0.6$, the left halves show the flow field, and the right halves show the droplet morphology. The result indicates that at $n=0.5$, a roll jet is generated without an ejecta sheet. A roll jet is a jet structure that curls at the beginning, rolls as it develops, and pushes the surrounding fluid outward (See Fig.\ S4 in Supplementary Material for a schematic diagram). The phenomenon of the roll jet was also observed during the impact of a Newtonian droplet on a liquid pool in the numerical simulation by Agbaglah et al.\ \cite{Agbaglah2015VortexAndJet}, in which the critical Capillary number for the formation of the roll jet was found to be $\text{Ca} \equiv {\text{We}}/{\text{Re}}<0.2$. However, the Capillary number in our case is $\text{Ca} \equiv {\text{We}}/{\text{Re}}=1.11$ (i.e., $\text{We}=219.91$, $\text{Re}=198.69$), which is much larger than the critical $\text{Ca}$ of the roll jet for Newtonian droplets. This difference can also be explained by the shear-thinning property of the fluid. A smaller value of $n$ indicates a stronger shear-thinning effect. As $n$ decreases, the effective viscosity at the point of the jet formation decreases, and the local viscous effect becomes less prominent. Since the formation of the roll jet relies on a small Capillary number $\text{Ca}<0.2$, i.e., a small ratio between the viscous force and the surface tension force, this condition becomes easier to be satisfied at the point of jet formation for the shear-thinning fluid. Therefore, the roll jet can occur for shear-thinning fluids even the Capillary number is much larger than the critical value for Newtonian fluids.

\subsection{ Effect of Od number }\label{sec:sec044}

\begin{figure*}
  \centerline{\includegraphics[width=1.7\columnwidth]{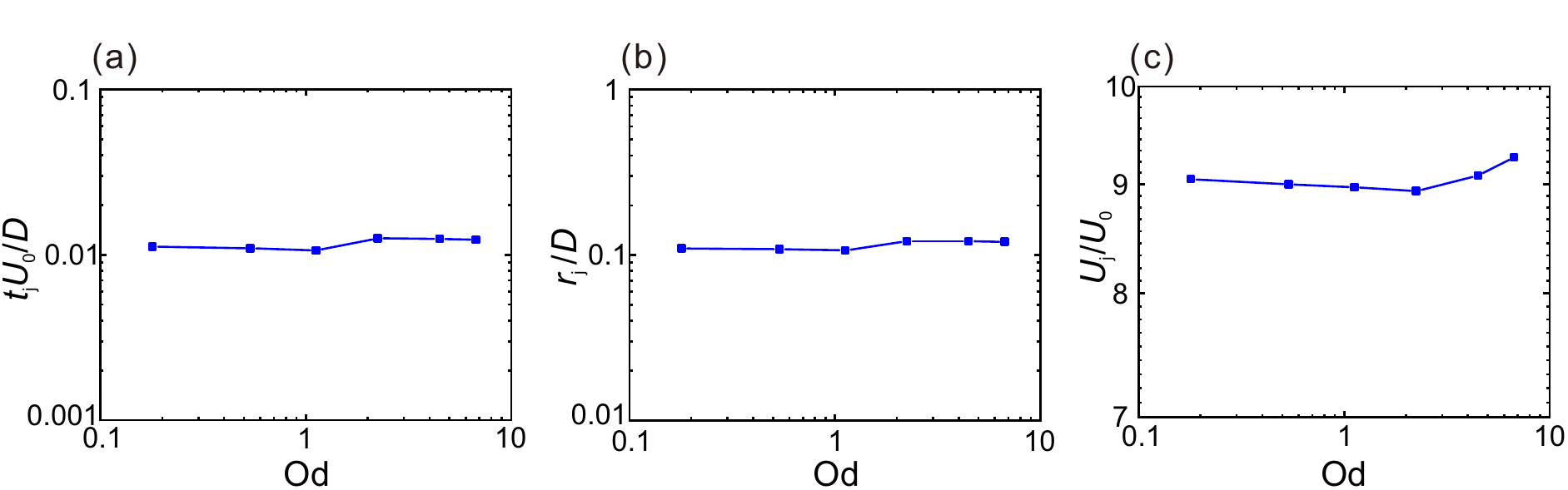}}%
  \caption{(a) The dimensionless time of ejecta sheet emergence ${{{t}_{\text{j}}}{{U}_{0}}}/{D}$, (b) the dimensionless spreading radius of ejecta sheet emergence ${{{r}_{\text{j}}}}/{D}$, and (c) the dimensionless velocity of ejecta sheet emergence ${{{U}_{\text{j}}}}/{{{U}_{0}}}$ as a function of the $\text{Od}$ number (i.e., $\text{Od}$ from 0 to 6.72). The Reynolds number is $\text{Re}=198.69$. The Weber number is $\text{We}=219.91$. }
\label{fig:fig10}
\end{figure*}

To analyze the influence of yield stress on the impact process, we fix other parameters and only change the yield stress ${{\tau }_{\text{c}}}$ from 0 to $150\ \text{Pa}\cdot \text{s}$ (i.e., $\text{Od}$ from 0 to 6.72) to explore its effect on the dimensionless time ${{{t}_{\text{j}}}{{U}_{0}}}/{D}$, the dimensionless spreading radius ${{{r}_{\text{j}}}}/{D}$, and the dimensionless velocity ${{{U}_{\text{j}}}}/{{{U}_{0}}}$ of ejecta sheet emergence. As shown in Fig.\ \ref{fig:fig10}, the time, the radius, and the velocity of the ejecta sheet emergence are hardly affected by the yield stress. Therefore, the yield stress is not the main factor affecting the emergence of the ejecta sheet. This is because the local shear rate at the point of ejecta sheet emergence is very high, corresponding to large local shear stress which is much larger than the yield stress. Therefore, the role of the yield stress ${{\tau }_{\text{c}}}$ on the emergence of the ejecta sheet can be negligible.

\begin{figure*}
  \centerline{\includegraphics[width=1.7\columnwidth]{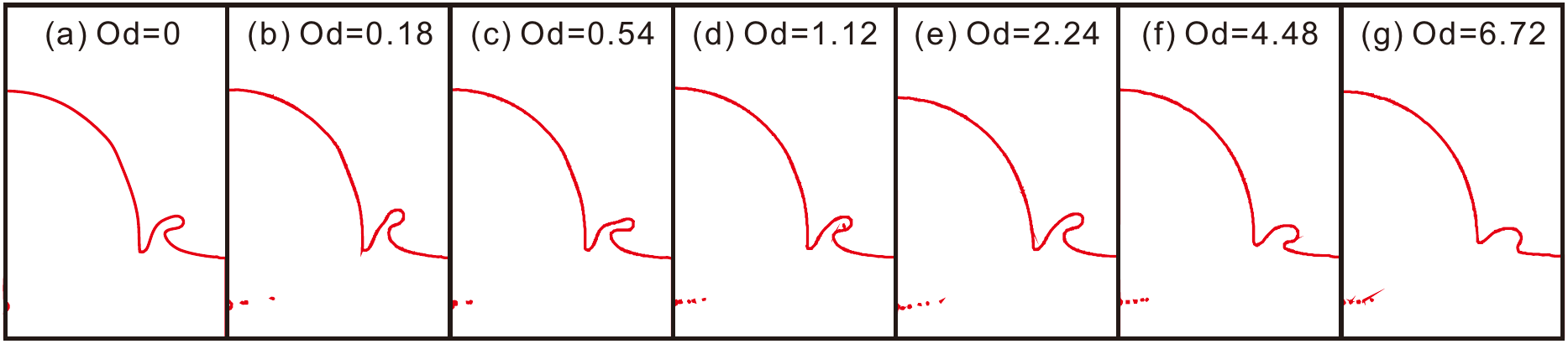}}%
  \caption{Shapes of the ejecta sheet for different $\text{Od}$ numbers at the same instant. The Reynolds number is $\text{Re}=198.69$; the Weber number is $\text{We}=219.91$; the dimensionless time is ${t{{U}_{0}}}/{D}=0.36$.}
\label{fig:fig11}
\end{figure*}

Even though the yield stress does not affect the time, the radius, and the velocity of the ejecta sheet emergence, it does affect on the shape of the ejecta sheet, particularly the thickness. The thickness of the ejecta sheet increases with the increase in the yield stress, as shown in Fig.\ \ref{fig:fig11}. According to the study by Josserand et al.\ \cite{Josserand2003DropletSplash}, the ejecta sheet thickness ${{e}_{\text{j}}}\left( t \right)$ is determined by the viscous boundary layer that is formed with the jet. For a shear-thinning fluid with a yield stress, the ejecta sheet thickness is affected by the effective rheology, i.e., ${{e}_{\text{j}}}\left( t \right)\sim {{t}^{1/2}}{{\rho }_{\text{L}}}^{{-1}/{2}}{{\left[ K{{\left( {D}/{{{U}_{0}}} \right)}^{\left( 1-n \right)}}+{{\tau }_{\text{c}}}\left( {D}/{{{U}_{0}}} \right) \right]}^{{1}/{2}}}$ (the derivation will be discussed in Sec.\ \ref{sec:sec045}). When the yield stress term ${{\tau }_{\text{c}}}\left( {D}/{{{U}_{0}}} \right)$ is much smaller than the viscosity term $K{{\left( {D}/{{{U}_{0}}} \right)}^{\left( 1-n \right)}}$, the yield stress term ${{\tau }_{\text{c}}}\left( {D}/{{{U}_{0}}} \right)$ can be ignored. In this condition, the ejecta sheet thickness is mainly affected by the consistency index $K$ and the flow index $n$. However, if the yield stress continues to increase, the yield stress term in the bracket ${{\tau }_{\text{c}}}\left( {D}/{{{U}_{0}}} \right)$ can no longer be ignored. Therefore, as $\text{Od}$ increases, the ejecta sheet thickness does not increase significantly at first. And as $\text{Od}$ increases further, the ejecta sheet thickness gradually increases, as shown in Fig.\ \ref{fig:fig11}.

\subsection{ Effect of Re number }\label{sec:sec045}

\begin{figure*}
  \centerline{\includegraphics[width=1.7\columnwidth]{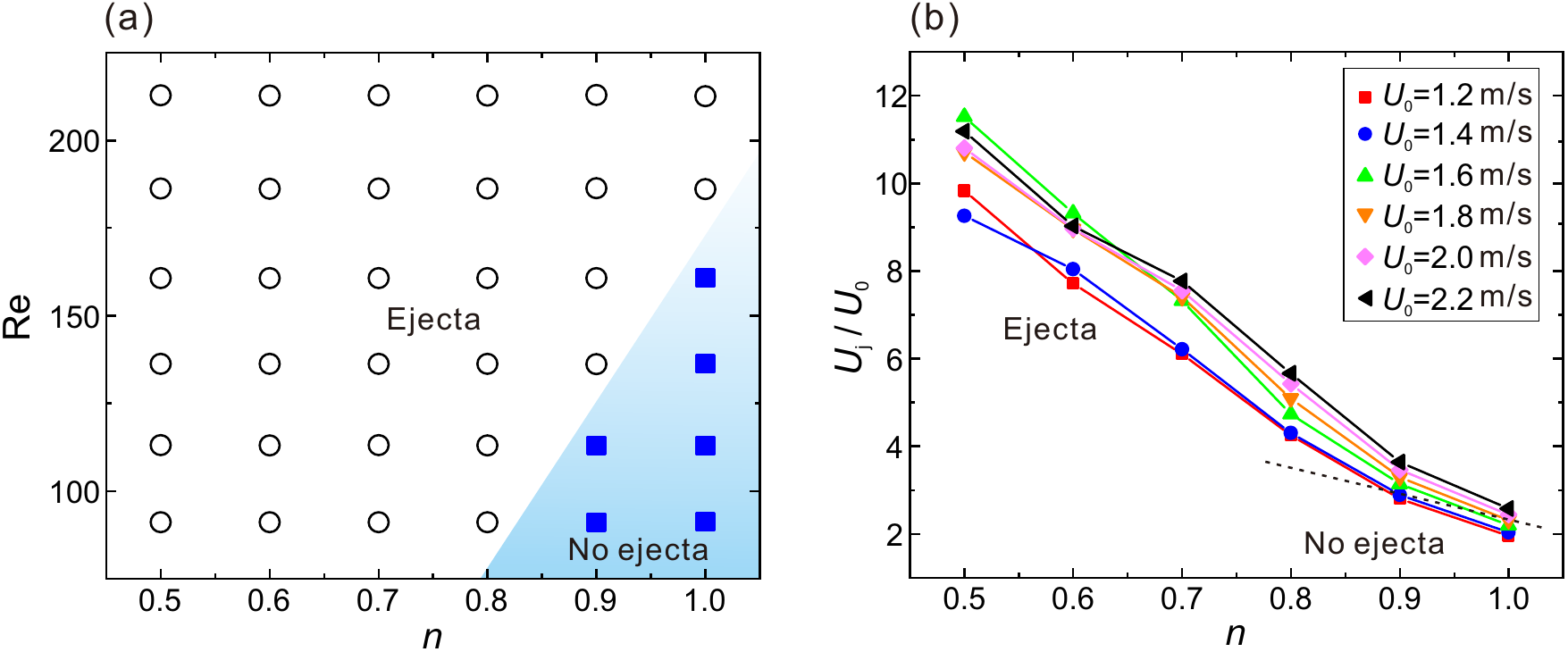}}%
  \caption{(a) Phase diagram of the ejecta as a function of $\text{Re}$ and $n$. The Reynolds number is between 91.05 and 212.84. The Weber number is between 72.14 and 242.48. (b) The dimensionless velocity of ejecta sheet emergence ${{{U}_{\text{j}}}}/{{{U}_{0}}}$ as a function of $n$.  }
\label{fig:fig12}
\end{figure*}

According to the definition of the Reynolds number in Eq.\ (\ref{eq:eq01}), many variables affect the Reynolds number, which quantifies the relative importance of the inertia and the viscous force at the characteristic shear rate ${{{U}_{0}}}/{D}$. Here, the Reynolds number is varied by changing the impact speed of the droplet ${{U}_{0}}$ and the consistency index $K$, separately, and their effects on the emergence of the ejecta sheet at different $n$ are analyzed in this section. In addition, the velocity of the ejecta sheet emergence was obtained by finding the peak in the curve of the maximum velocity, as discussed in Sec.\ \ref{sec:sec042}. Here the maximum is with respect to the spatial domain, and the peak in the curve is with respect to time. For the no-ejecta cases, there is still a peak in the curve of the maximum velocity, but the magnitude of the peak velocity is not enough to generate the ejecta sheet.

By varying the impact speed of the droplet ${{U}_{0}}$ and the flow index $n$, a regime map for the formation of the ejecta sheet is produced in the $\text{Re}-n$ space, as shown in Fig.\ \ref{fig:fig12}a. It can be shown that the ejecta sheet does not appear at large $n$ and small $\text{Re}$. The variation in the dimensionless velocity of the ejecta sheet emergence is plotted in Fig.\ \ref{fig:fig12}b, and it shows that the velocity of the ejecta sheet emergence decreases as $n$ increases. This trend is also consistent for different speeds of the droplet impact. According to the critical condition of splashing \cite{Josserand2003DropletSplash, Thoraval2012DropletSplash} ${{\text{We}}^{{1}/{2}}}{{\text{Re}}^{{1}/{4}}}\ge {{K}_{\text{c}}}$ where ${{K}_{\text{c}}}$ is a constant, the viscosity is important to the ejecta sheet emergence. When $n$ approaches 1, the effective viscosity is almost uniform in the whole domain. The local effective viscosity at the contact point between the droplet and the pool (i.e., the point of ejecta sheet formation) is large. Therefore, when the impact speed is low and $n$ approaches 1, an ejecta sheet is difficult to form.

\begin{figure*}
  \centerline{\includegraphics[width=1.7\columnwidth]{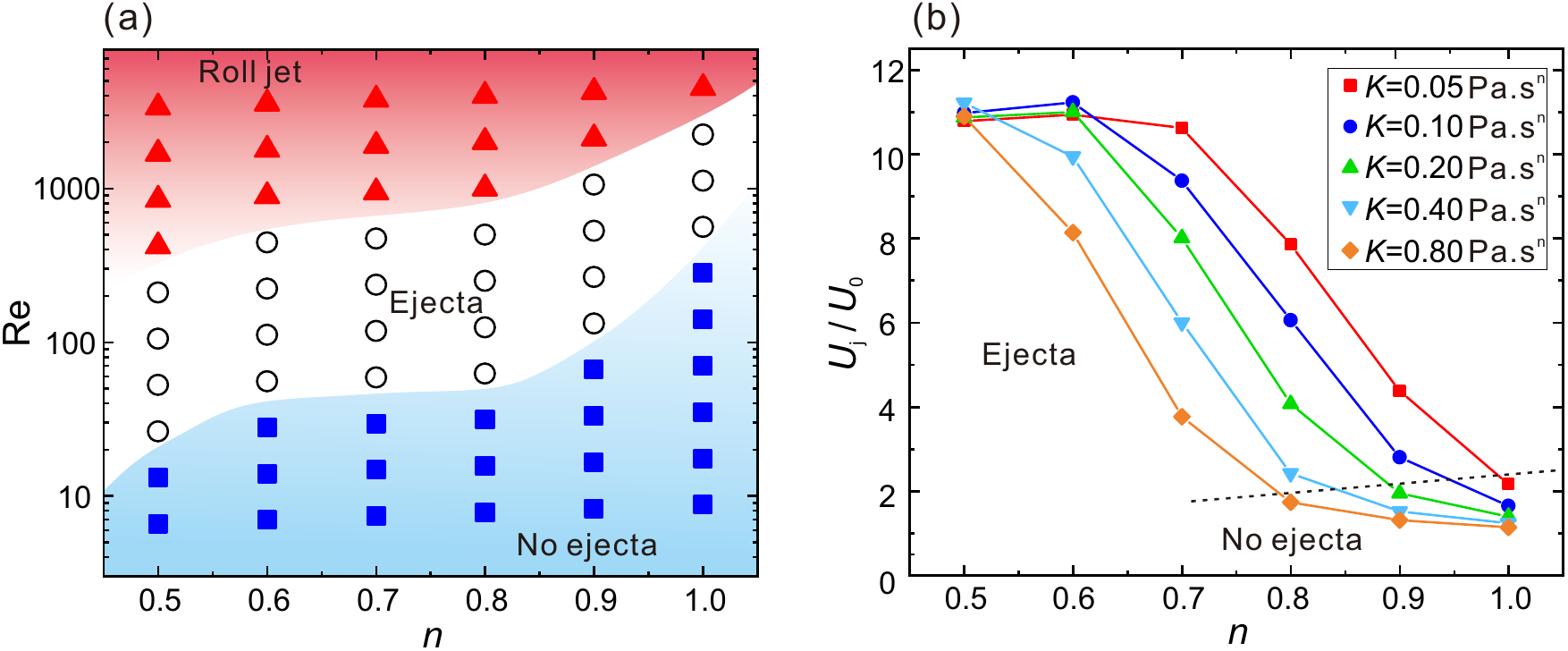}}%
  \caption{(a) Phase diagram of the ejecta as a function of $\text{Re}$ and $n$. The Reynolds number is between 8.77 and 3366.70. The Weber number is $\text{We}=200.40$. (b) The dimensionless velocity of ejecta sheet emergence ${{{U}_{\text{j}}}}/{{{U}_{0}}}$ as a function of $n$.}
\label{fig:fig13}
\end{figure*}

We also vary the Reynolds number by changing the consistency index $K$, and produce a regime map in the $\text{Re}-n$ space, as shown in Fig.\ \ref{fig:fig13}. Three phenomenas occur as $n$ and $K$ vary, namely no ejecta, ejecta, and roll jet. Ejecta will not be generated when $\text{Re}$ is small and $n$ is large, and this is mainly because the viscous dissipation is very strong, and the large velocity gradient that is necessary for vortex formation and jet formation cannot be achieved. As the Reynolds number increases, the viscous dissipation effect reduces, and the ejecta sheet will be generated. When $\text{Re}$ is large and $n$ is small, the viscous effect is very weak, and the roll jet will be generated.

The variation in the dimensionless velocity of the ejecta sheet emergence is plotted in Fig.\ \ref{fig:fig13}b. The velocity of the ejecta sheet emergence decreases as $n$ increases, but the variation is nonlinear. The dimensionless velocity of the ejecta sheet emergence is not sensitive to $K$ when $n$ approaches 0.5 or 1. It can also be explained by the shear-thinning property. As $n$ decreases, the shear-thinning effect increases, and the local effective viscosity at the point of the ejecta sheet emergence is small. The effect of $K$ on the local effective viscosity is weaker than that of $n$. Therefore, the dimensionless velocity of the ejecta sheet emergence is not sensitive to $K$ when $K$ is small.

In contrast, when $n$ approaches 1, the fluid property approximates the Newtonian fluid, and the effective viscosity is almost uniform in the whole domain. Therefore, the local effective viscosity at the contact point between the droplet and the pool (i.e., the point of ejecta sheet formation) is large, and an ejecta sheet is difficult to form. As a consequence, the local dimensionless velocity is small, and no ejecta sheet is produced.

\subsection{Theoretical analysis of the ejecta sheet}\label{sec:sec046}

Regarding the emergence of the ejecta sheet during the impact of Newtonian droplets, a model was proposed by Josserand et al.\ \cite{Josserand2016DropletSplash, Josserand2003DropletSplash}, considering the mass conservation of incompressible steady flow. The volume flow rate of the immersed liquid is equal to the volume flow rate of the thin sheet produced
\begin{equation}\label{eq:eq08}
  {{Q}_{\text{m}}}\left( t \right)\sim 2\pi {{r}_{\text{g}}}\left( t \right){{e}_{\text{j}}}\left( t \right){{U}_{\text{j}}}.
\end{equation}
where ${{Q}_{\text{m}}}\left( t \right)$ is the volume flow rate of the immersed liquid at a time $t$ without considering the droplet deformation, and ${{e}_{\text{j}}}\left( t \right)$ is the thickness of the ejecta sheet. The volume flow rate of the immersed liquid ${{Q}_{\text{m}}}\left( t \right)$ can be estimated from the droplet impact speed and the geometrical radius ${{r}_{\text{g}}}\left( t \right)$ discussed in Sec.\ \ref{sec:sec042}
\begin{equation}\label{eq:eq09}
  {{Q}_{\text{m}}}\left( t \right)\sim \pi {{r}_{\text{g}}}\left( t \right)^{2}{{U}_{0}}.
\end{equation}
The ejecta sheet thickness ${{e}_{\text{j}}}\left( t \right)$ depends on the thickness of the viscous boundary layer on the free surface \cite{Josserand2016DropletSplash, Josserand2003DropletSplash}
\begin{equation}\label{eq:eq10}
  	{{e}_{\text{j}}}\left( t \right)\sim {{\left( {{\nu }_{\text{L}}}t \right)}^{1/2}},
\end{equation}
where ${{\nu }_{\text{L}}}\equiv{{{\mu }_{\text{L}}}}/{{{\rho }_{\text{L}}}}$ is the kinematic viscosity of the droplet. Based on Eqs.\ (\ref{eq:eq08})--(\ref{eq:eq10}), the velocity of the ejecta sheet emergence can be obtained. Here, the Reynolds number uses the original definition $\text{Re} \equiv {{{\rho }_{\text{L}}}D{{U}_{0}}}/{{{\mu }_{\text{L}}}}$.
\begin{equation}\label{eq:eq11}
  	{{{U}_{\text{j}}}}/{{{U}_{0}}}\sim {\frac{1}{2}{\text{Re}}^{{1}/{2}}}.
\end{equation}
The ejecta sheet appears only when its velocity is larger than the geometrical velocity ${{U}_{\text{j}}}>{{u}_{\text{g}}}\left( t \right)$, where ${{u}_{\text{g}}}\left( t \right)$ is the geometrical velocity defined as ${{u}_{\text{g}}}\left( t \right)\equiv\frac{1}{2}{{\left( {D{{U}_{0}}}/{t} \right)}^{1/2}}$. Otherwise, if ${{U}_{\text{j}}}<{{u}_{\text{g}}}\left( t \right)$, the ejecta sheet will be overrun by the falling droplet. Therefore, ${{U}_{\text{j}}}\sim {{u}_{\text{g}}}\left( t \right)$ is the critical condition of the ejecta sheet formation, we can have
\begin{equation}\label{eq:eq12}
  {\frac{1}{2}{\text{Re}}^{{1}/{2}}}{{U}_{0}}\sim \frac{1}{2}{{\left( \frac{D{{U}_{0}}}{{{t}_{\text{g}}}} \right)}^{1/2}}.
\end{equation}
Rearranging Eq.\ (\ref{eq:eq12}), we have
\begin{equation}\label{eq:eq13}
  {{t}_{\text{j}}}\sim {{t}_{\text{g}}}\sim \frac{1}{\text{Re}}\frac{D}{{{U}_{0}}}.
\end{equation}
Then, the similarity relation of the emergence time of the ejecta sheet can be finally obtained
\begin{equation}\label{eq:eq14}
  {{{t}_{\text{j}}}{{U}_{\text{0}}}}/{D}\sim {{\text{Re}}^{-1}}.
\end{equation}

In this study of shear-thinning fluids with a yield stress, the Herschel--Bulkley model $\tau ={{\tau }_{\text{c}}}+K{{\dot{\gamma }}^{n}}$ can be used to calculate the effective viscosity ${{\mu }_{\text{L}}}={{{\tau }_{\text{c}}}}/{{\dot{\gamma }}}+K{{\dot{\gamma }}^{n-1}}$. By substituting it into Eq.\ (\ref{eq:eq10}), we can obtain the ejecta sheet thickness for the non-Newtonian fluid
\begin{equation}\label{eq:eq15}
  {{e}_{\text{j}}}\left( t \right)\sim {{t}^{1/2}}{{\rho }_{\text{L}}}^{{-1}/{2}}{{\left[ K{{\left( {D}/{{{U}_{0}}} \right)}^{\left( 1-n \right)}}+{{\tau }_{\text{c}}}\left( {D}/{{{U}_{0}}} \right) \right]}^{{1}/{2}}}.
\end{equation}
The emergence velocity of the ejecta sheet can be obtained by substitute the effective viscosity into Eq.\ (\ref{eq:eq11})
\begin{equation}\label{eq:eq16}
  {{{U}_{\text{j}}}}/{{{U}_{0}}}\sim \frac{1}{2}{{{\text{Re}}^{{1}/{2}}}}/{{{\left( 1+\text{Od} \right)}^{{1}/{2}}}}.
\end{equation}
The emergence time of the ejecta sheet can be obtained by substituting the effective viscosity into Eq.\ (\ref{eq:eq14})
\begin{equation}\label{eq:eq17}
  	{{{t}_{\text{j}}}{{U}_{\text{0}}}}/{D}\sim \left( 1+\text{Od} \right){{\text{Re}}^{-1}}.
\end{equation}

\begin{figure*}
  \centerline{\includegraphics[width=1.4\columnwidth]{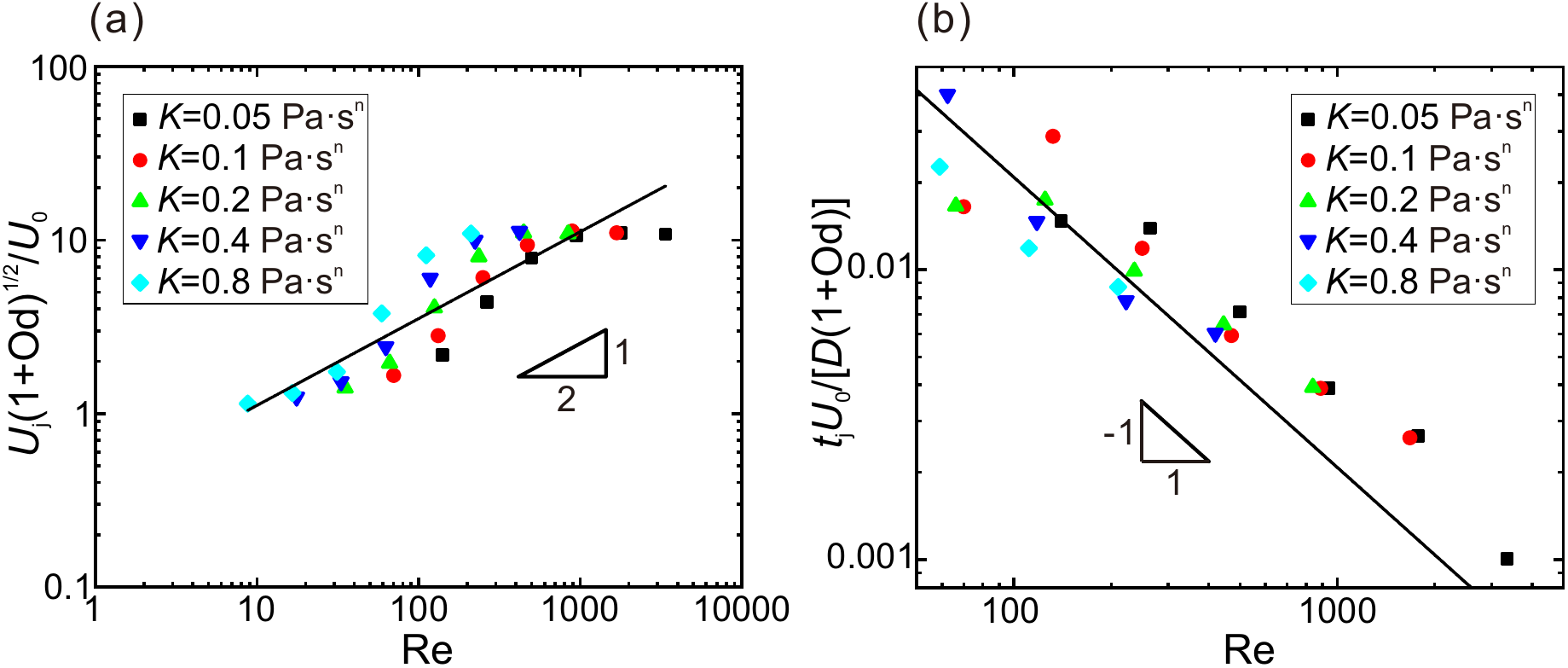}}%
  \caption{(a) Log-log plot of ${{{U}_{\text{j}}}{{\left( 1+\text{Od} \right)}^{{1}/{2}}}}/{{{U}_{0}}}$ against $\text{Re}$, changing $K$ from $0.05\ \text{Pa}\cdot {{\text{s}}^{\text{n}}}$ to $0.8\ \text{Pa}\cdot {{\text{s}}^{\text{n}}}$ and $n$ from 0.5 to 1. The Oldroyd number is $\text{Od}=0$. The solid line is ${{{U}_{\text{j}}}{{\left( 1+\text{Od} \right)}^{{1}/{2}}}}/{{{U}_{0}}}=0.32{{\text{Re}}^{{1}/{2}}}$ fitted from our simulated data. (b) Log-log plot of ${{{t}_{\text{j}}}{{U}_{\text{0}}}}/{\left[D\left( 1+\text{Od} \right) \right]}$ against $\text{Re}$, changing $K$ from $0.05\ \text{Pa}\cdot {{\text{s}}^{\text{n}}}$ to $0.8\ \text{Pa}\cdot {{\text{s}}^{\text{n}}}$ and $n$ from 0.5 to 1. The Oldroyd number is $\text{Od}=8.35$. The solid line is ${{{t}_{\text{j}}}{{U}_{\text{0}}}}/{\left[D\left( 1+\text{Od} \right) \right]}=2.08{{\text{Re}}^{-1}}$ fitted from our simulated data. }
\label{fig:fig14}
\end{figure*}

To verify the above analysis, we compare it with the numerical data obtained by changing $K$ from 0.05 to $0.8\ \text{Pa}\cdot {{\text{s}}^{\text{n}}}$ and changing $n$ from 0.5 to 1, as shown in Fig.\ \ref{fig:fig14}(a). Since the $\text{Od}$ number has negligible influence on the emergence of the ejecta sheet (as shown in Fig.\ \ref{fig:fig10} and discussed in Sec.\ \ref{sec:sec044}), $\text{Od}$ remains unchanged during our simulation. Because the viscous dissipation and the droplet deformation are neglected in the model, the theoretical emergence velocity of the ejecta sheet will be larger than the numerical data. In contrast, the theoretical time of the ejecta sheet emergence will be lower than the numerical data. We can find that the prefactor of the fitting solid line ${{{U}_{\text{j}}}{{\left( 1+\text{Od} \right)}^{{1}/{2}}}}/{{{U}_{0}}}=0.32{{\text{Re}}^{{1}/{2}}}$ in Fig.\ \ref{fig:fig14}(a) is reasonably described by the constant 0.32, which is in the reasonable range relative to the prefactor ${1/2}$ of the theoretical emergence velocity of the ejecta sheet in Eq.\ (\ref{eq:eq16}). 
Regarding the time of the ejecta sheet emergence, the numerical data are  plotted according to the scaling of Eq.\ (\ref{eq:eq17}), as shown in Fig.\ \ref{fig:fig14}(b).
The comparison shows that the prefactor 2.08 of the fitting solid line is in the reasonable range relative to the prefactor 1 of the theory of the emergence time of the ejecta sheet.

\section{ Conclusions }\label{sec:sec05}
In this study, we consider the splashing process of shear-thinning droplets with a yield stress, and focus on the effects of fluid rheology on the ejecta sheet emergence. The formation and the propagation of the ejecta sheet are analyzed, and the velocity, the radius, and the time of the ejecta sheet emergence are used to quantify the ejecta sheet. Regarding the ejecta sheet formation, the ejecta sheet becomes easier to form as the flow index $n$ reduces, confirming that the shear-thinning effect can promote the ejecta sheet formation. Large yield stress can effectively affect the thickness of the ejecta sheet but can hardly change the ability of the ejecta sheet emergence. As the ejecta sheet develops, the dimensionless spreading radius is found to collapse into a geometrical radius ${{r}_{\text{g}}}\left( t \right)=\sqrt{D{{U}_{0}}t}$ as predicted according to the mass conservation, due to that the inertia force is the dominant factor of this process. The scaling of the ejecta sheet for shear-thinning fluid with yield stress is also verified by comparing it with numerical data. The results of this study are not only useful for unveiling the mechanism of splashing dynamics during the impact of droplets, but are also helpful for understanding other behaviors of non-Newtonian droplets, such as deformation, breakup, and coalescence. The effects of other non-Newtonian properties also deserve systematic studies, such as shear-thickening, thixotropic, and viscoelastic properties, which exist in a wide range of applications in chemical engineering, material synthesis, bioengineering, etc.

\section*{Acknowledgements}
This work is supported by the National Natural Science Foundation of China (Grant nos.\ 51676137 and 52176083).

\section*{Data Availability Statement}
The data that support the findings of this study are available from the corresponding author upon reasonable request.

\section*{References}
\bibliography{nonNewtonianDropletImpact}
\end{document}